\documentclass[times,twocolumn,final]{elsarticle}

\usepackage{lineno,hyperref}

\journal{Journal of \LaTeX\ Templates}

\usepackage{times}
\usepackage{epsfig}
\usepackage{graphicx}
\usepackage{amsmath}
\usepackage{amssymb}
\usepackage{bbm}
\usepackage{xcolor}
\usepackage{amsmath} 
\usepackage{multirow,booktabs}
\usepackage{mathtools}
\usepackage{soul}
\usepackage{xcolor}

\mathchardef\mhyphen="2D 
\newcommand{\cross}[1][1pt]{\ooalign{%
  \rule[.8ex]{1.5ex}{#1}\cr
  \hss\rule{#1}{.7em}\hss\cr}}
  
\definecolor{red}{rgb}{0.234, 0.6835, 0.808}

\renewcommand{\hl}[1]{#1}

\begin{document}

\begin{frontmatter}

\title{Region-wise Loss for Biomedical Image Segmentation}

\author{Juan Miguel Valverde, Jussi Tohka\corref{mycorrespondingauthor}}
\address{A.I. Virtanen Institute for Molecular Sciences, University of Eastern Finland, 70150 Kuopio, Finland}
\cortext[mycorrespondingauthor]{Corresponding author}
\ead{\{juanmiguel.valverde,jussi.tohka\}@uef.fi}

\begin{abstract}
We propose Region-wise (RW) loss for biomedical image segmentation.
Region-wise loss is versatile, can simultaneously account for class imbalance and pixel importance, and it can be easily implemented as the pixel-wise multiplication between the softmax output and a RW map.
We show that, under the proposed RW loss framework, certain loss functions, such as Active Contour and Boundary loss, can be reformulated similarly with appropriate RW maps, thus revealing their underlying similarities and a new perspective to understand these loss functions.
We investigate the observed optimization instability caused by certain RW maps, such as Boundary loss distance maps, and we introduce a mathematically-grounded principle to avoid such instability.
This principle provides excellent adaptability to any dataset and practically ensures convergence without extra regularization terms or optimization tricks.
Following this principle, we propose a simple version of boundary distance maps called rectified Region-wise (RRW) maps that, as we demonstrate in our experiments, achieve state-of-the-art performance with similar or better Dice coefficients and Hausdorff distances than Dice, Focal, weighted Cross entropy, and Boundary losses in three distinct segmentation tasks.
We quantify the optimization instability provided by Boundary loss distance maps, and we empirically show that our RRW maps are stable to optimize.
The code to run all our experiments is publicly available at: https://github.com/jmlipman/RegionWiseLoss.

\end{abstract}

\begin{keyword}
Deep Learning \sep Segmentation \sep Medical Imaging \sep Loss Function
\end{keyword}
\end{frontmatter}


\section{Introduction}

Image segmentation is a typical pre-processing step required for quantitative image analyses in biomedical applications.
Accurate segmentation is crucial to measure, for instance, brain tumor size, which can determine the radiation dose administered to patients during radiotherapy \cite{kim2017prognostic}.
Segmenting images manually is time consuming and subjective, as evidence by the large disagreement between manual segmentations from different annotators (see e.g., \cite{valverde2020ratlesnetv2}).
Thus, there is a great interest in developing reliable tools to segment medical images automatically \cite{liu2021review}.

Convolutional neural networks (ConvNets) have shown excellent performance in multiple segmentation tasks, leading to their consideration in clinical applications, such as pathological feature segmentation in optical coherence tomography of retina \cite{wilson2021validation}, glioblastoma segmentation \cite{perkuhn2018clinical}, and head and neck segmentation for radiotherapy \cite{nikolov2021clinically}.
In medical image segmentation, accounting for class imbalance and pixel importance is critical to produce accurate segmentations.
By pixel importance, we refer to the phenomenon in which misclassifications severity depends on the location of the misclassifications.
For instance, in tumor segmentation, misclassifications far from the tumor boundaries can be more severe than near the boundaries. 
To consider class imbalance or pixel importance, minimizing multiple terms or loss functions simultaneously is a common strategy.
Taghanaki et al. \cite{taghanaki2019combo} combined Cross entropy loss with Dice loss to tackle class imbalance.
Gerl et al. \cite{gerl2020distance} combined Cross entropy with a domain-specific penalty term.
Peng et al. \cite{peng2021dgfau} combined Dice loss and Focal loss functions in a DenseNet-based architecture, and Kamran et al. \cite{kamran2021rv} combined Hinge loss and mean squared error to increase the reconstruction accuracy in a generative adversarial network.
However, minimizing multiple loss functions or terms requires extra hyper-parameters to balance the individual contribution of each loss function, as loss functions yield values in different ranges, which, in turn, affect the gradients during back-propagation.
As a consequence, costly hyper-parameter tuning becomes necessary.
Furthermore, entangling multiple losses makes unclear whether a specific loss function \textit{alone} or the extra terms were responsible for tackling, e.g., class imbalance.
In this paper, we propose Region-wise (RW) loss, a general framework capable of jointly considering class imbalance and pixel importance without additional hyper-parameters or loss functions.

We show that several widely used loss functions, including Active contour (AC) \cite{chen2019learning}, Hausdorff distance (HD) \cite{karimi2019reducing} and Boundary \cite{kervadec2018boundary} losses, can be reformulated as RW losses, revealing their underlying similarities.
Furthermore, Boundary and HD losses are reportedly unstable to optimize and, as explained in \cite{karimi2019reducing,kervadec2018boundary}, it is necessary to combine them with Dice loss to circumvent such optimization instability.
However, it is unclear if specifically Dice loss should be paired with Boundary or HD losses to achieve stable optimization and, more importantly, the exact cause of this optimization instability has remained unknown.
Kervadec et al. \cite{kervadec2018boundary} hypothesized that the optimization of boundary distance maps occasionally fails because empty foregrounds (all zeroes in the foreground class) produce small gradients. This hypothesis describes a typical class imbalance scenario where the majority of the pixels outweigh foreground pixels and, consequently, the few foreground pixels contribute little to the gradients. In this paper, we explain, utilizing the RW loss framework, why these loss functions are unstable to optimize, and we introduce a mathematically-grounded principle that leads to optimization stability.

\begin{figure*}[t]
\begin{center}
   \includegraphics[width=0.98\linewidth]{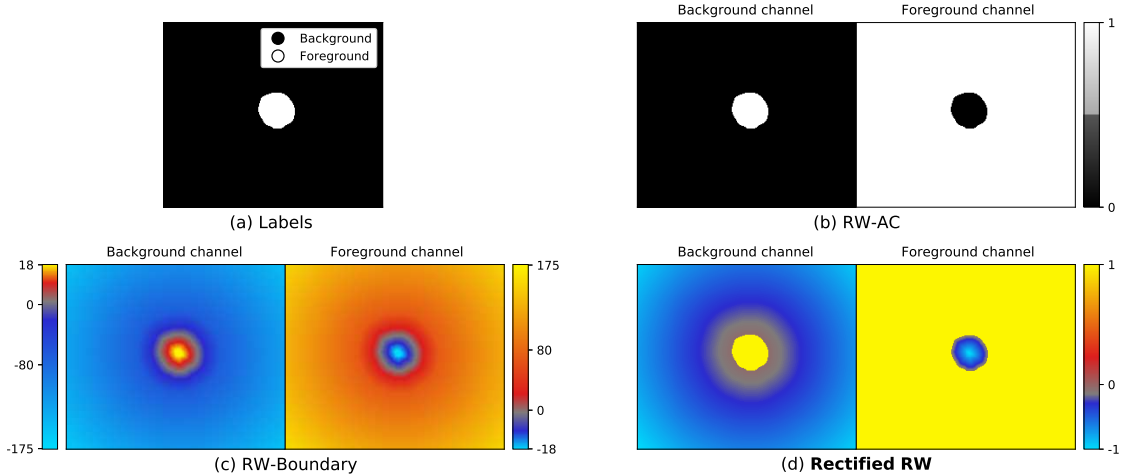}
\end{center}
   \caption{Representative visualization of the labels of an image from a two-class segmentation task (a) and its corresponding RW maps (b-c-d). (a) Labels. (b) RW Active contour map. (c) RW Boundary distances map. (d) Our proposed normalized RRW map, with normalized negative distances in the foreground and ones in the background.}
\label{fig:maps}
\end{figure*}

Our contributions are the following:

\begin{itemize}
\item We present RW loss, a loss function that is unique in its ability to simultaneously account for class imbalance and pixel importance. Further, we show that certain important loss functions (AC \cite{chen2019learning}, HD \cite{karimi2019reducing} and Boundary \cite{kervadec2018boundary} losses) can be reformulated.
\item We analyze RW loss and provide theoretical and empirical insight of the cause of the observed---yet not understood---optimization instability and we introduce a principle to fix such instability. We apply this principle to derive rectified Region-wise (RRW) maps (Fig. \ref{fig:maps}, (d)) that require no extra loss functions, regularization terms, or optimization tricks to ensure optimization stability.
\item We show that RW loss with our proposed RRW maps achieved state-of-the-art performance, producing similar or better Dice coefficients and Hausdorff distances than Dice, Focal, weighted Cross entropy and Boundary losses on three large biomedical datasets \cite{bakas2018identifying,bernard2018deep,1904.00445}.
\item We compared previous strategies to circumvent optimization instability with Boundary loss, and we show that, regardless of the strategy, our RRW maps led to less fluctuations during the optimization and provided convergence faster.
\end{itemize}

\section{Related work}
\paragraph{Segmentation with class imbalance}
Class imbalance occurs when pixels from a certain class outnumber other classes, a common scenario in medical image segmentation (e.g., tumor vs. non-tumor).
One simple method to tackle class imbalance is to assign higher weights to foreground classes, as in weighted Cross entropy loss.
Focal loss \cite{lin2017focal} modifies Cross entropy to reduce the effect of high softmax probability values and it accentuates pixels with smaller softmax probabilities.
Several loss functions have been inspired by widely used segmentation performance metrics, including Dice loss \cite{milletari2016v,sudre2017generalised}, Tversky loss \cite{salehi2017tversky}, Jaccard loss \cite{bertels2019optimizing}, Lovász-Softmax loss \cite{berman2018lovasz}, and HD loss \cite{karimi2019reducing}. 
Active contour (AC) loss \cite{chen2019learning}, inspired by the classical Chan-Vese formulation of active contour models \cite{chen2019learning}, also focuses on the foreground classes.
AC loss divides the images into foreground and background and, as we detail in the next section, it maximizes the predicted probability values of true positives and negatives while considering the length of the predicted boundary.

\paragraph{Boundary distance maps} In medical image segmentation, misclassification severity can depend on pixel/voxel location.
Frequently, misclassifications near the segmentation boundaries are less severe than those farther apart, since these boundaries can be ambiguous to denote.
Hausdorff distance quantifies this misclassification severity by measuring the largest segmentation error.
Boundary loss \cite{kervadec2018boundary}, similarly to HD loss \cite{karimi2019reducing}, penalizes heavily misclassifications far from the segmentation boundaries with boundary distance maps (Fig \ref{fig:maps} (c)).
Wang et al. \cite{wang2021active} employed these boundary distance maps to gradually align ground truth's boundaries with the prediction's.
Caliva et al. \cite{caliva2019distance} reversed boundary distance map values by assigning more importance to areas near the segmentation boundaries.
Gerl et al. \cite{gerl2020distance} incorporated prior knowledge about the epidermis shape through a distance-based smoothness penalty term.
Notably, the aforementioned works employed distance-based loss functions that were unstable to optimize and, to circumvent such instability, extra loss functions were also optimized, as suggested in \cite{kervadec2018boundary}.
Furthermore, certain strategies to optimize loss functions, such as in Boundary loss \cite{kervadec2018boundary}, update hyper-parameters after every epoch and, consequently, are heavily affected by the training set size, as smaller training sets perform less iterations per epoch, leading to faster hyper-parameter updates.
Motivated by the reliance of these loss functions on creative strategies and extra hyper-parameters, we investigated their instability and, based on our analyses, propose our RRW maps that require no extra loss functions or hyper-parameters while utilizing boundary distance maps.

\section{Region-wise loss}
\subsection{Definition and gradients}

Let $\boldsymbol{x} \in \mathbb{R}^N$ be a vectorized image and $\boldsymbol{Y} = (y)_{ik} \in \{0,1\}^{N \times K}$ be the one-hot encoded matrix of its labels where $N$ and $K$ are the number of pixels and classes, respectively.
Importantly, the background is also considered as a class, and every pixel $i$ belongs to one of the $K$ classes.
Let $\boldsymbol{\Phi} = (\phi)_{ik} = [\boldsymbol{\phi}_1, \ldots, \boldsymbol{\phi}_N]$ denote the unnormalized prediction of a ConvNet (i.e., logits) and $\boldsymbol{\hat{Y}} = (\hat{y})_{ik} = [\boldsymbol{\hat{y}}_1, \ldots, \boldsymbol{\hat{y}}_N]$ its softmax normalized values, widely used in semantic segmentation due to the validity of such probability values. We define RW loss as the sum of the element-wise multiplication between the softmax probability values and a RW map $\boldsymbol{Z} = (z)_{ik} = [\boldsymbol{z}_1, \ldots, \boldsymbol{z}_N]$. Formally:

\begin{equation} \label{eq:rwdefinition}
L_{RW} =  \sum_{i=1}^N \boldsymbol{\hat{y}}_i^{\top} \boldsymbol{z}_i = \sum_{i=1}^N \sigma(\boldsymbol{\phi}_i)^{\top} \boldsymbol{z}_i
\end{equation}
where $\boldsymbol{\phi}_i$, $\boldsymbol{\hat{y}}_i$ and $\boldsymbol{z}_i$ are $K$-length column vectors and $\sigma$ is the softmax function.
In definition (\ref{eq:rwdefinition}), we make two assumptions that simplify the notation and cause no loss of generality: 1) we have omitted the normalization by $(NK)^{-1}$, which we use in our implementation; 2) following \cite{chen2019learning,kervadec2018boundary,sudre2017generalised}, we restrict the notation to a single training image with the assumption that the loss function with $J$ training images is computed as the average value of a single image loss function.
This formulation (Eq. (\ref{eq:rwdefinition})) extends \cite{kervadec2018boundary} and it is not limited to a specific RW map.
RW maps are computed as $\boldsymbol{Z} = f(\boldsymbol{Y})$ where $f: \{0,1\}^{N \times K} \to \mathbb{R}^{N \times K}$.
Importantly, $\boldsymbol{Z}$ depends on the ground truth and is independent of the network's parameters. Figure \ref{fig:maps} (b-c-d) shows three examples of RW maps.
Region-wise loss yields the following gradients with respect to the unnormalized prediction of a ConvNet $\boldsymbol{\Phi}$:

\begin{equation} \label{eq:rwderivative}
\begin{split}
\frac{\partial L_{RW}}{\partial \phi_{ik}} & = \frac{\partial L_{RW}}{\partial \boldsymbol{\hat{y}}_{i}} \frac{\partial \boldsymbol{\hat{y}}_{i}}{\partial \phi_{ik}} = z_{ik} \hat{y}_{ik} (1 - \hat{y}_{ik}) + \sum_{l = 1; l \neq k}^K z_{il} (-\hat{y}_{ik} \hat{y}_{il}) \\
& = z_{ik} \hat{y}_{ik} - z_{ik} \hat{y}_{ik} \hat{y}_{ik} - \sum_{l = 1; l \neq k}^K z_{il} \hat{y}_{ik} \hat{y}_{il} \\
& = z_{ik} \hat{y}_{ik} - \hat{y}_{ik} \sum_{l=1}^K z_{il} \hat{y}_{il} = \hat{y}_{ik} \sum_{l = 1; l \neq k}^K \hat{y}_{il} (z_{ik} - z_{il})
\end{split}
\end{equation}
that multiply every other gradient during the optimization with back-propagation.
Details about the full derivation can be found in Section 1 of the Supplementary Material.
Note that $\boldsymbol{z}_i$ with equal values at all the components (e.g., $\boldsymbol{z}_i=[t, t, t, t]$) will yield zero gradients, which should be avoided when defining $\boldsymbol{Z}$.

\subsection{Properties} \label{sec:properties}
Region-wise loss is versatile, and other loss functions can be reformulated as a RW loss.
Region-wise loss generalizes Boundary loss \cite{kervadec2018boundary} that \hl{can be seen as utilizing} RW maps defined as the Euclidean distance transform to segmentation boundaries.
We refer to these RW maps as RW-Boundary maps (Fig. \ref{fig:maps} (c)), and they are defined as

\begin{equation} \label{eq:rwboundarymaps}
z_{ik}=
    \begin{cases}
        -{||i - b_{ik}}||_2 & \text{if } i \in \Omega_k \\
        {||i - b_{ik}}||_2 & \text{otherwise}
    \end{cases},
\end{equation}
where $\Omega_k$ is the set of pixels/voxels in class $k$ in the ground-truth, and $b_{ik}$ is the closest ground-truth boundary pixel to $i$ in class $k$. Likewise, one-sided HD loss \cite{karimi2019reducing} that uses distance maps as Boundary loss can also be reformulated as RW loss with a slight modification.
Considering binary segmentation into foreground and background, active contour (AC) loss $L_{AC}$ \cite{chen2019learning}, inspired by active contour models, combines the regions inside and outside the ground truth foreground $\Omega$ with the $length$ of the mask perimeter (or surface, in 3D images): $L_{AC} = length + \lambda (R_{in} + R_{out})$, where $R_{in} = \sum_{i \in \Omega} \hat{y}_{i1}$ \hl{are the foreground pixels misclassified as background (i.e., false negatives)}, $R_{out} = \sum_{i \notin \Omega}\hat{y}_{i2}$ \hl{are the background pixels misclassified as foreground (i.e., false positives)}, and $1$ and $2$ refer to the background and foreground classes, respectively.
By defining a RW map as $\boldsymbol{z}_i = [1,0]^T$ if $i \in \Omega$ and $\boldsymbol{z}_i = [0,1]^T$ if $i \notin \Omega$, or, equivalently, $\boldsymbol{Z} = \boldsymbol{1} - \boldsymbol{Y}$, where $\boldsymbol{1}$ is the $N \times K$ matrix of ones, we get $L_{AC} = length + \lambda L_{RW \mhyphen AC}$ (Fig. \ref{fig:maps} (a-b) \hl{illustrates $\boldsymbol{Y}$ and $\boldsymbol{Z}$, respectively}). Intuitively, multiplying the softmax probability values by $\boldsymbol{Z}$ penalizes background pixels (i.e., the false positives and negatives) in proportion to their corresponding softmax values; and because softmax probabilities sum to one, minimizing the softmax values of background pixels equates to maximizing the softmax values of the foreground area.
This reformulation of AC loss into RW loss reveals its a priori non-trivial similarity with another loss function \cite{cao2020boundary} that proposed non-distance maps defined as $\boldsymbol{z}_i = [-\beta,\alpha]^T$ if $i \in \Omega$ and $\boldsymbol{z}_i = [\alpha,-\beta]^T$ if $i \notin \Omega$.
In other words, under our RW loss framework, AC loss is the zero-one normalized version of \cite{cao2020boundary}.
\hl{All the steps to reformulate these loss functions into RW loss can be found in Section 2 of the Supplementary Material.}

Region-wise loss allows penalizing each predicted pixel in a certain manner based on its class \textit{and} location.
Location-based penalty can be implemented as in Eq. (\ref{eq:rwboundarymaps}), and class-based penalty can be further added with high values in $z_{ik}$ for all $i \in \Omega_m$, and for all $k \neq m$, where $m$ is the class to be emphasized.
This flexibility contrasts with other loss functions such as Cross entropy. Cross entropy can be modified to tackle class imbalance (weighted Cross entropy), and it can also be adjusted to incorporate pixel-wise-specific information \cite{ronneberger2015u}. However, incorporating weights similar to RW maps into the Cross entropy loss to account for class imbalance and pixel importance simultaneously as $L_{PWCE} = - \sum_{i,k}^{N,K} z_{ik} y_{ik} \log{\hat{y}_{ik}}$ does not guarantee that the gradients will utilize all RW map values.
\hl{Since its derivative is $\frac{\partial L_{PWCE}}{\partial \phi_{ik}} = -z_{ik} y_{ik} + \hat{y}_{ik} \sum_l z_{il} y_{il}$ (derivation analogous to Eq. (}\ref{eq:rwderivative}), see Section 1 of the Supplementary Material) and $\boldsymbol{y}_i$ is a one-hot encoded vector, only $z_{ik}$ s.t. $y_{ik} = 1$ will contribute to the gradient, consequently, $K-1$ values will remain unused.
Thus, tackling class imbalance and pixel importance with Cross entropy without any extra term is not trivial.

\subsection{Rectified Region-wise maps (RRW)}
\hl{We derive a simple mathematically-grounded principle to correct RW maps for ensuring optimization stability.}
Specifically, any RW map can be rectified by enforcing $z_{ik} = z_{il} $ in Eq. (\ref{eq:rwderivative}) for all pixels $i \notin \Omega_k \cup \Omega_l$, i.e., by setting the $K-1$ ``background" components of each $\boldsymbol{z}_i$ vector to have an equal value.
\hl{If $i \in \Omega_k, z_{ik} \in \mathbb{R}_{\geq 0}, z_{il} \in \mathbb{R}_{\leq 0}$ for all $l \neq k$, then $\frac{\partial L_{R W}}{\partial \phi_{i k}} \leq 0$ and $\frac{\partial L_{R W}}{\partial \phi_{i l}} \geq 0$ that, as we describe in Section }\ref{sec:sourcesinstability}\hl{, is key for a stable optimization.}
\hl{Note that Boundary loss (Eq. (}\ref{eq:rwboundarymaps}\hl{)) in segmentation tasks with more than two foreground classes does not follow this principle whereas AC loss }\cite{chen2019learning}\hl{ that reported no optimization instability follows, inadvertently, this principle.}

We applied the same principle to RW-Boundary maps (Eq. (\ref{eq:rwboundarymaps})) and we normalized foreground values between $[-1, 0]$, yielding our \hl{RRW} maps $\boldsymbol{Z}$:

\begin{equation} \label{eq:rrwmaps}
z_{ik}=
    \begin{cases}
        \frac{-{||i - b_{ik}}||_2}{max_{j \in \Omega_k} {||j - b_{jk}}||_2} & \text{if } i \in \Omega_k \\
        1              & \text{otherwise}
    \end{cases}.
\end{equation}
In other words, foreground values are normalized negative Euclidean distances to the boundaries (Fig. \ref{fig:maps} (d)).
This simple approach is hyper-parameter-free and yields favorable gradients in typical optimization scenarios.
Furthermore, these values that cancel out $z_{ik}, z_{il}$ also represent pixel importance and affect the computed loss.
\hl{Thus, these values can be used to increase/decrease the loss in pixels from specific classes to handle class imbalance}.
\hl{For instance, for softmax values $\boldsymbol{\hat{y}}_i = [0.1, 0.1, 0.2, 0.6]$ and RW map values $\boldsymbol{z}_i = [0, 0, 0, -0.6]$, the gradients are $[0.04, 0.04, 0.07, -0.14]$ and the loss becomes $-0.36$ (before normalizing by multiplying by $(NK)^{-1}$).
In contrast, for more important pixels, setting $\boldsymbol{z}_i = [2, 2, 2, -0.6]$ yields larger gradients $[0.16, 0.16, 0.31, -0.62]$ and a loss of $0.44$.}
Since too large gradients can excessively alter ConvNets' parameters during the optimization, we suggest to normalize these RW maps to facilitate training stability and convergence under the most common optimization hyper-parameters choices.
\hl{Additionally, RW maps can be computed during the optimization to apply transfer learning by decreasing/increasing the contribution of certain images to the loss }\cite{pan2009survey}\hl{ (e.g. by decreasing/increasing $z_{ik}$ for all $i \notin \Omega_k$)}.

\subsection{Source of optimization instability} \label{sec:sourcesinstability}

\hl{During the optimization of ConvNets with softmax at the end of their architecture, loss functions, such as Cross entropy, produce gradient vectors $\frac{\partial L}{\partial \boldsymbol{\phi}_{i}}$ with one negative component and $K-1$ positive components.}
\hl{Since the ground truth is a one-hot encoded vector, one gradient component, say $k$, must be negative and the other gradient components must be positive for each pixel $i$. This leads to the corresponding softmax value $\hat{y}_{i k}$ to increase and others $\hat{y}_{il}$, $l \neq k$, to decrease.}
In contrast, certain loss functions that can be reformulated as \hl{RW} loss, including Boundary loss \cite{kervadec2018boundary}, use RW maps that produce partial derivatives with incorrect signs. These incorrect signs arise due to the interconnections $z_{ik}$ and $z_{il}$ (Eq. (\ref{eq:rwderivative})) in segmentation tasks of more than two classes. 
\hl{Specifically, if there exists $\boldsymbol{z}_i$ for which two or more partial derivatives $\frac{\partial L}{\partial \phi_{i k}} < 0$, then $\boldsymbol{Z}$ is inadequately designed and multiple softmax values $\boldsymbol{\hat{y}}_i$ can increase at the same time, causing optimization instability.}

\begin{figure}[t]
\begin{center}
   \includegraphics[width=0.98\linewidth]{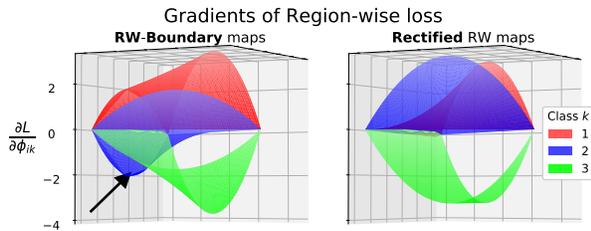}
\end{center}
   \caption{\hl{Gradients of RW loss with respect to a RW-Boundary map value ($\boldsymbol{z}_i= [12, 4, -3]$) and a RRW map value ($\boldsymbol{z}_i= [10, 10, -3]$). The horizontal plane is the 2-simplex that represents all possible softmax vectors $\boldsymbol{\hat{y}}_i$. The black arrow indicates the presence of undesirable negative values in the gradient vector components corresponding to class 2.}}
\label{fig:gradientsrectifiedunrectified}
\end{figure}

Figure \ref{fig:gradientsrectifiedunrectified} shows the gradients produced by a particular $\boldsymbol{z}_i$ from RW-Boundary (left) and our RRW map (right) for all possible softmax vectors in a three-class segmentation task. As the true label is $\boldsymbol{y}_i = [0, 0, 1]$, gradient vectors are expected to contain one \hl{negative component corresponding to} class 3 (green) and two \hl{positive components corresponding to} classes 1 and 2 (red and blue). However, the simulation depicted in Figure \ref{fig:gradientsrectifiedunrectified} (left, black arrow) shows that \hl{more than half of the gradient vectors produced by RW-Boundary map contained two negative components}, increasing two softmax values simultaneously. In contrast, our RRW map yielded \hl{gradient components} with the correct sign across all softmax vectors.

\begin{figure*}[t]
\begin{center}
   \includegraphics[width=0.99\linewidth]{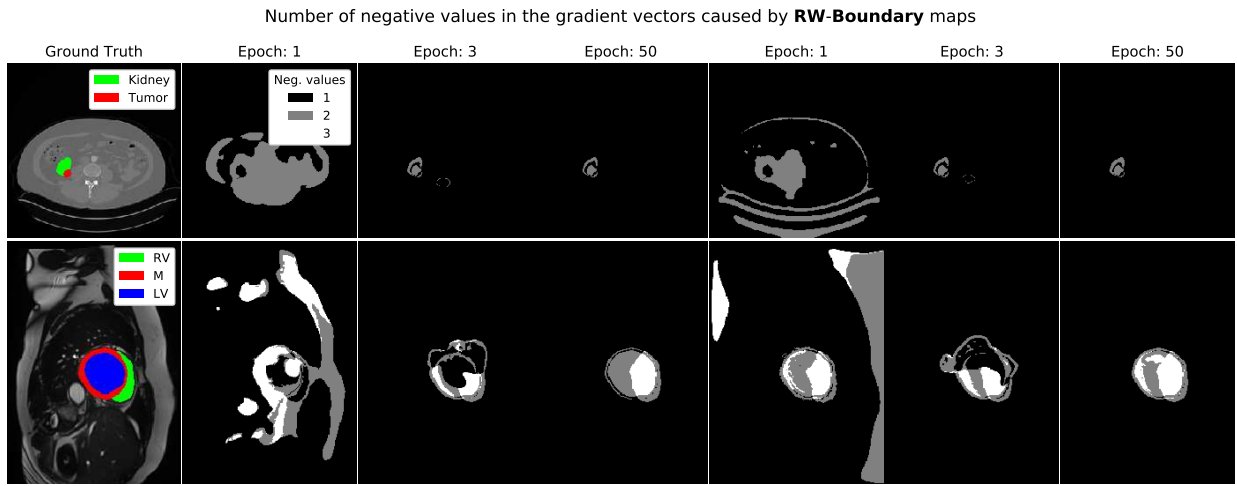}
\end{center}
   \caption{Number of negative values in the gradient vectors when optimizing with RW-Boundary maps. Top row: KiTS19 dataset. Bottom row: ACDC17 dataset. First column: Ground truth. Columns 2-4 and 5-7 show two independent runs at different optimization time-points.}
\label{fig:gradientsepochs}
\end{figure*}

Figure \ref{fig:gradientsepochs} illustrates the number of negative \hl{components in the gradient vectors} in images from two datasets used in this study (KiTS19 (top) and ACDC17 (bottom)) at three different optimization time-points in two independent training instances. Contrary to the expected single negative \hl{component} per softmax vector, RW-Boundary maps produced gradient vectors $\frac{\partial L_{RW}}{\partial \boldsymbol{\phi}_i}$ with several negative \hl{components}, guiding their corresponding softmax vectors to undesirably increase in multiple directions. The regions with \hl{gradient components} with incorrect signs (Fig. \ref{fig:gradientsepochs}, negative values 2 and 3) start large and progressively shrink. However, these regions do not completely disappear, hindering the optimization.

\begin{figure}[t]
\begin{center}
   \includegraphics[width=0.98\linewidth]{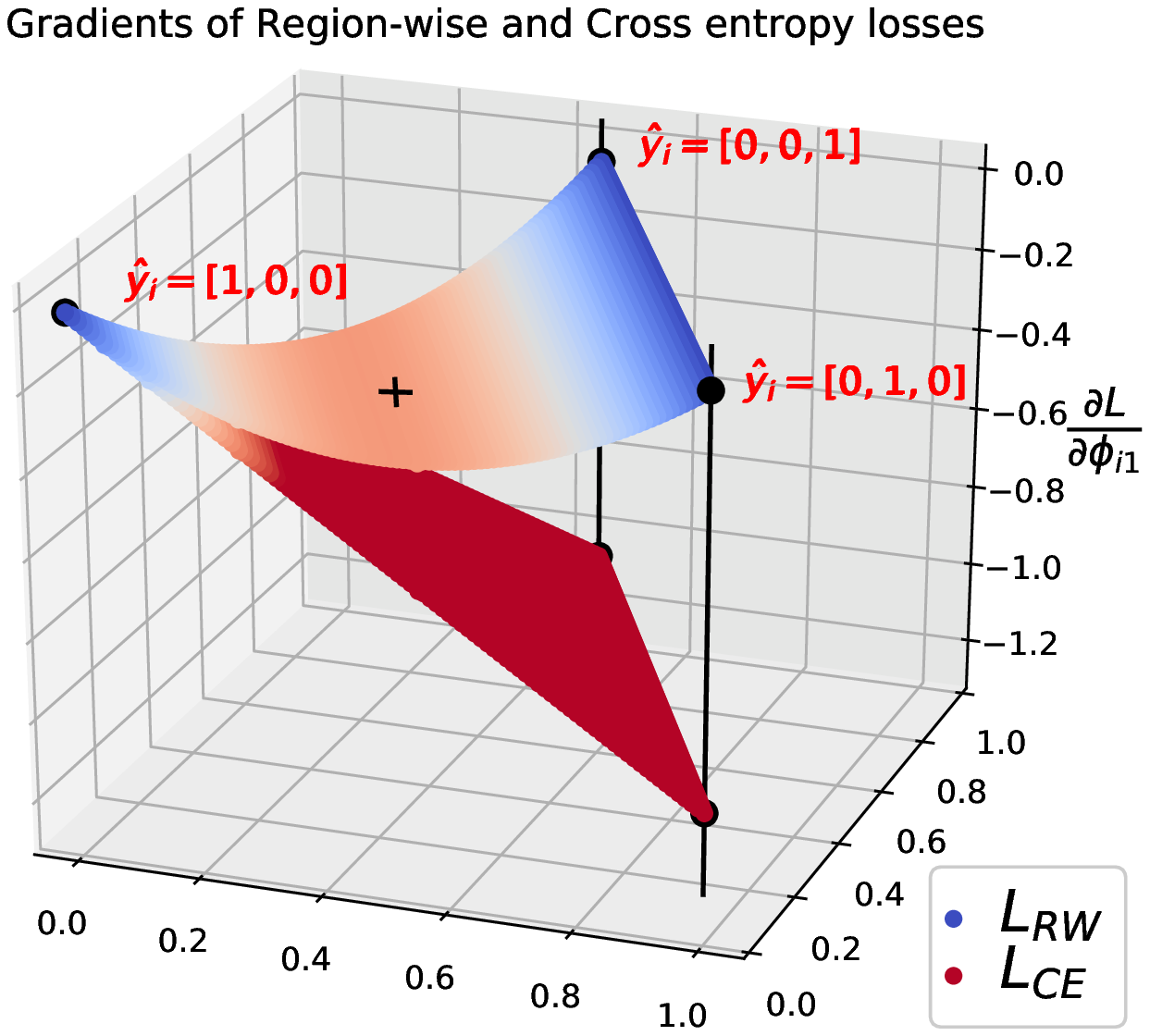}
\end{center}
   \caption{\hl{Gradients of RW and Cross entropy losses with respect to $\phi_{i1}$ in a three-class segmentation problem. The horizontal plane is the 2-simplex that represents all possible softmax vectors $\boldsymbol{\hat{y}}_i$.} \cross[.4pt]: saddle point where $\frac{\partial L_{RW}}{\partial \phi_{i1}} = -0.25$}
\label{fig:gradients}
\end{figure}

In addition to the instability caused by certain RW maps, \hl{RW loss gradient components $\frac{\partial L_{RW}}{\partial \phi_{ik}}$ do not always increase in proportion to the error}, and these gradients can occasionally be too small. 
Despite these two issues related to small and non-linear gradients can be circumvented with suitable step size, \hl{and since these issues arise with any RW map}, we study them to provide a deeper understanding of RW loss.
Following Eq. (\ref{eq:rwderivative}), the gradients in a two-class segmentation task are:

\begin{equation}
\begin{split}
& \frac{\partial L_{RW}}{\partial \phi_{i1}} = z_{i1} \hat{y}_{i1} - \hat{y}_{i1}^2 z_{i1} - \hat{y}_{i1} z_{i2} \hat{y}_{i2} \\
& = \hat{y}_{i1} ( \hat{y}_{i2} (z_{i1} - z_{i2} )) = \hat{y}_{i1} (1 -\hat{y}_{i1} ) (z_{i1} - z_{i2}).
\end{split}
\end{equation}
Since $(z_{i1} - z_{i2})$ is constant, softmax values $\hat{y}_{i1} = t$ and $\hat{y}_{i1} = 1 - t$ with $t \in [0,1]$ produce the same gradient component.
\hl{Moreover, such gradient component becomes proportionally larger to $\hat{y}_{i1}$ with $\hat{y}_{i1} \in [0, 0.5]$, and proportionally smaller with $\hat{y}_{i1} \in [0.5, 1]$}.
This property is not limited to two-class segmentation. Figure \ref{fig:gradients} illustrates RW and Cross entropy loss gradients with respect to $\phi_{i1}$ for all possible combinations of softmax values in a three-class segmentation task with $\boldsymbol{y}_i = \boldsymbol{1} - \boldsymbol{z}_i = [1, 0, 0]$. Cross entropy gradients (Figure \ref{fig:gradients}, lower triangle) are correlated to the error: the farther $\boldsymbol{\hat{y}}_i$ is from the true label the gradient becomes more negative. In contrast, RW loss gradients (Figure \ref{fig:gradients}, upper triangle) behave similarly until the saddle point at the center of the triangle where $\frac{\partial L_{RW}}{\partial \phi_{i1}} = -0.25$ and gradients start increasing.
Additionally, since the softmax value $\hat{y}_{ik}$ (Eq. (\ref{eq:rwderivative})) multiplies every other term of the summation, small softmax values yield small gradient \hl{components}, potentially hindering the optimization at the beginning when the softmax values associated to the true labels can be small.
\hl{On the other hand, note that these gradients (Eq. (}\ref{eq:rwderivative}\hl{), Fig. }\ref{fig:gradients}\hl{), important to understand optimization stability, are only at one specific layer, and that all CNN layers affect the output}; in practice, adaptive optimization algorithms, such as Adam \cite{Kingma2014AdamAM}, provide sufficiently good step sizes, facilitating the optimization.
In fact, in our experiments \hl{(Sections }\ref{sec:evaluation}\hl{-}\ref{sec:regionwisecomparison}\hl{ )}, we demonstrate that optimizing RW loss is not only feasible but it also achieves state-of-the-art performance.

\section{Experiments} \label{sec:experiments}
We conducted the following experiments: 1) comparing RW loss and our hyper-parameter-free \hl{RRW} maps with \hl{four} other widely used loss functions on three large biomedical datasets; 2) evaluating RW loss with RRW maps on ACDC17 competition's test set; 3) studying the convergence frequency provided by RW-Boundary and our RRW maps; and 4) evaluating multiple strategies to achieve convergence with unrectified RW maps and our RRW maps.

\begin{table}
\begin{center}
\begin{tabular}{l|ccc}
Dataset & C & Train/Val Images & Description \\
\hline
ACDC17 \cite{bernard2018deep} & 3 & 100/100 (\hl{3D}) & Cardiac MRI \\
BraTS18 \cite{bakas2018identifying} & 3 & 105/105 (\hl{3D}) & Brain MRI \\
KiTS19 \cite{1904.00445} & 2 & 105/105 (\hl{3D}) & Kidney CT \\
\end{tabular}
\end{center}
\caption{Dataset summary. C is the number of foreground classes.} \label{table:data}
\end{table}

\paragraph{Datasets}
We conducted our experiments on Automated Cardiac Diagnosis Challenge 2017 (ACDC17) \cite{bernard2018deep}, Multimodal Brain Tumor Segmentation Challenge 2018 (BraTS18) \cite{bakas2018identifying} and Kidney Tumor Challenge 2019 (KiTS19) \cite{1904.00445} training sets (see Table \ref{table:data} for a brief overview).
ACDC17 dataset comprised cardiac magnetic resonance images (MRIs) of 100 patients acquired with different equipment.
For each subject, two 3D-image frames were manually segmented into background, right ventricle (RV), myocardium (M) and left ventricle (LV).
BraTS18 dataset included the 210 high-grade glioma patients, and the images were segmented into background, edema (ED), enhancing tumor core (ET) and necrotic and non-enhancing tumor core (NCR\_NET).
KiTS19 dataset included 210 patients, and images were segmented into background, kidney and kidney tumor. Additionally, we resized KiTS19 image slices to 256x256 to reduce the computational costs.
\hl{The three datasets were divided to training and validation sets, both containing 50\% of the subjects, at the subject level so that the training subject set and the validation subject set were mutually exclusive. We standardized all the images to have zero-mean unit-variance.}
Note that this split is considerably more challenging than in the segmentation competitions since our training sets are much smaller.

\paragraph{Model and training}
We utilized nnUNet \cite{isensee2021nnu}, a self-configuring UNet \cite{ronneberger2015u} that tailors certain optimization settings and architectural components (e.g., number of filters, batch size, image crop size) to each dataset.
nnUNet has achieved state-of-the-art results on several segmentation datasets \cite{isensee2021nnu} and, since it is based on UNet, it portrays a typical segmentation approach in biomedical image segmentation.
We utilized the official implementation of nnUNet\footnote{https://github.com/MIC-DKFZ/nnUNet}; we added the evaluated loss functions (RW, Dice \cite{milletari2016v}, Focal \cite{lin2017focal}, \hl{weighted Cross entropy}, Boundary loss \cite{kervadec2018boundary}), and we employed Adam optimizer \cite{Kingma2014AdamAM}, as in \cite{kervadec2018boundary}, with a learning rate starting at \hl{$10^{-4}$}.
\hl{The choice of this learning rate, which is smaller than in nnUNet }\cite{isensee2021nnu}\hl{, was based on our preliminary experiments.}
\hl{We computed the RRW maps on the 3D images, and we optimized 2D models since it has been shown that training on 2D slices of 3D anisotropic images provides more accurate models} \cite{isensee2021nnu}.
\hl{We randomly initialized nnUNet following He et al.} \cite{he2015delving} \hl{(i.e., weights $W \sim \mathcal{N}(0, 2/n)$ where $n$ is the number of input neurons)}, and we trained it for 300 epochs in ACDC17 dataset, and 100 epochs in BraTS18 and KiTS19 datasets.
\hl{The remaining optimization settings of nnUNet, including polynomial learning rate decrease and deep supervision }\cite{lee2015deeply}\hl{, were the same as in the original implementation and  across all our experiments.}
\hl{Furthermore, nnUNet's automatic configuration was conducted on our local machine with an NVidia GeForce GTX 1080 Ti with 11 GB of memory, once per dataset, ensuring that nnUNet's self-configuration components that depend on the available computing resources remained fixed in all our experiments.}
All the experiments were implemented in Pytorch \cite{paszke2019pytorch}, and the complete code to run all our experiments and scripts to prepare the data can be found at: https://github.com/jmlipman/RegionWiseLoss.

\paragraph{Evaluation} \hl{The evaluation was done on the 3D images} with Dice coefficient \cite{dice1945measures} and Hausdorff distance \cite{rote1991computing}. Dice coefficient quantifies the similarity between the predictions and their ground truth

\begin{equation}
    Dice(A, B) = \frac{2|A \cap B|}{|A| + |B|},
\end{equation}
where $A$ and $B$ are the segmentation masks. Hausdorff distance measures the largest segmentation error

\begin{equation}
HD(A, B)=\max \left\{\max _{a \in \partial A} \min _{b \in \partial B}|b-a|, \max _{b \in \partial B} \min _{a \in \partial A}|a-b|\right\},
\end{equation}
where $\partial A$ and $\partial B$ are the boundary voxels of $A$ and $B$, respectively. Particularly, Hausdorff distance is crucial to demonstrate that pixel importance is considered during the optimization.
Since certain object boundaries can be ambiguous to delineate due to low image contrast---typical in medical images, pixels far from these boundaries are critical, and small Hausdorff distances are desired.
\hl{Hausdorff distances were reported in millimeters (mm.).}
We averaged Dice coefficients and Hausdorff distances across three independent runs and, in the experiments where we evaluated convergence separately (Section \ref{sec:facilitatingconvergence}), we rerun models that failed to converge, avoiding convergence problems to interfere with performance values.

\subsection{Evaluation on three large datasets} \label{sec:evaluation}

We evaluated RW loss with our hyper-parameter-free RRW maps ($L_{RRW}$, Eq. (\ref{eq:rrwmaps})) on ACDC17, BraTS18 and KiTS19 datasets.
We compared our approach with Boundary loss, which combines RW-Boundary maps (Eq. (\ref{eq:rwboundarymaps})) and Dice loss, and, as our RRW maps, also considers pixel importance via distance maps.
We optimized Boundary loss following the original training setting \cite{kervadec2018boundary}, i.e., by decreasing $\alpha$ after every epoch from $1$ to $0.01$ in: $L_{Dice+Boundary} = \alpha L_{Dice} + (1 - \alpha) L_{RW}$.
We also evaluated Dice and Focal losses, two state-of-the-art loss functions that account for class imbalance.
We optimized the non-$\alpha$-balanced Focal loss function (i.e., $\alpha = 1$ \hl{in $L_{Focal} = - \alpha (1 - \hat{y})^{\gamma} log(\hat{y}) $}), and we set $\gamma = 2$, as the original study reported that this choice yielded the best results \cite{lin2017focal}.
\hl{Additionally, we evaluated Cross entropy weighted with our RRW maps as indicated in Section }\ref{sec:properties}\hl{ since, despite several RW map values not contributing to the gradient, Cross entropy's gradients are proportional to the error, unlike RW loss' gradients (see Fig. }\ref{fig:gradients}\hl{).}
Tables \ref{table:acdc17}, \ref{table:brats18} and \ref{table:kits19} list the per-class Dice coefficients and HDs, and the proportion of voxels per class.
\hl{We tested the hypothesis whether performance measure values (Dice coefficient, Hausdorff distance) in the 3D images in the validation set differed between our RRW loss and the other studied losses. The hypothesis was tested using a paired two-sample permutation test using the mean-absolute difference of the performance measure between the two losses as the test statistic. The permutation distribution was approximated using 10000 permutations. We used the non-parametric permutation test instead of the parametric t-test as the performance measures cannot be assumed to be normally distributed. We considered $p$-values smaller than 0.05 as statistically significant.}

{\renewcommand{\arraystretch}{1.2}
\setlength{\tabcolsep}{8pt}
\begin{table*}
\begin{center}
\scriptsize{
\hspace*{-36pt}  \begin{tabular}{l|cc|cc|cc}
& \multicolumn{2}{c}{  RV (1.21 \%) } & \multicolumn{2}{c}{  M (1.3 \%) } & \multicolumn{2}{c}{  LV (1.27 \%) } \\
Loss \hl{function} & Dice & HD (mm.) & Dice & HD (mm.) & Dice & HD (mm.) \\
\hline
$L_{RRW}$ & \ 0.90 $\pm$ 0.09 & 14.58 $\pm$ 11.03 & 0.90 $\pm$ 0.03 & 9.01 $\pm$ 7.62 & 0.94 $\pm$ 0.05 & 7.53 $\pm$ 5.95 \\
\hl{$L_{WCE(RRW)}$} & \ 0.87 $\pm$ 0.09$^{*}$ & 16.23 $\pm$ 10.44 & 0.87 $\pm$ 0.04$^{*}$ & 8.85 $\pm$ 5.23 & 0.94 $\pm$ 0.05 & 7.43 $\pm$ 5.50 \\ 
$L_{Dice}$ & \ 0.87 $\pm$ 0.11$^{*}$ & 20.98 $\pm$ 15.34$^{*}$ & 0.89 $\pm$ 0.03$^{*}$ & 13.87 $\pm$ 21.83$^{*}$ & 0.94 $\pm$ 0.05 & 11.40 $\pm$ 19.29$^{*}$ \\ 
$L_{Focal}$ & \ 0.89 $\pm$ 0.09 & 14.89 $\pm$ 11.07 & 0.90 $\pm$ 0.03 & 9.90 $\pm$ 14.60 & 0.94 $\pm$ 0.05 & 8.44 $\pm$ 13.44 \\
$L_{Dice+Boundary}$ & \ 0.89 $\pm$ 0.09 & 16.55 $\pm$ 13.85 & 0.90 $\pm$ 0.04 & 8.41 $\pm$ 5.39 & 0.94 $\pm$ 0.06 & 8.36 $\pm$ 7.85 \\
\end{tabular}
}
\end{center}
\caption{Performance on ACDC17 dataset. \hl{Percentages in parenthesis indicate the proportion of voxels per class (background: 96.22 \%)}. Scores that were significantly different from $L_{RRW}$ were marked with * ($p$-value $< 0.05$).} \label{table:acdc17}
\end{table*}}

{\renewcommand{\arraystretch}{1.2}
\setlength{\tabcolsep}{8pt}

\begin{table*}
\begin{center}
\scriptsize{
\hspace*{-36pt} \begin{tabular}{l|cc|cc|cc}
& \multicolumn{2}{c}{  ET (0.26 \%) } & \multicolumn{2}{c}{ ED (0.66 \%) } & \multicolumn{2}{c}{  NCR\_NET (0.16 \%) } \\
Loss \hl{function} & Dice & HD (mm.) & Dice & HD (mm.) & Dice & HD (mm.) \\
\hline
$L_{RRW}$ & 0.80 $\pm$ 0.15 & 20.76 $\pm$ 19.32  & 0.78 $\pm$ 0.14 & 37.48 $\pm$ 18.56 & 0.61 $\pm$ 0.26 & 16.47 $\pm$ 12.64 \\
\hl{$L_{WCE(RRW)}$} & 0.80 $\pm$ 0.14 & 23.59 $\pm$ 18.75 & 0.76 $\pm$ 0.13$^{*}$ & 35.76 $\pm$ 18.83 & 0.61 $\pm$ 0.26 & 16.39 $\pm$ 11.99 \\
$L_{Dice}$ &  0.79 $\pm$ 0.16$^{*}$ & 77.49 $\pm$ 16.67$^{*}$ & 0.77 $\pm$ 0.14 & 74.22 $\pm$ 13.83$^{*}$ & 0.61 $\pm$ 0.25 & 61.48 $\pm$ 22.34$^{*}$ \\
$L_{Focal}$ & 0.80 $\pm$ 0.14 & 19.27 $\pm$ 18.16 & 0.78 $\pm$ 0.14 & 37.01 $\pm$ 19.25 & 0.61 $\pm$ 0.26 & 16.16 $\pm$ 14.63 \\
$L_{Dice+Boundary}$ & 0.80 $\pm$ 0.15 & 19.04 $\pm$ 15.58 & 0.78 $\pm$ 0.14 & 29.35 $\pm$ 18.80$^{*}$  & 0.60 $\pm$ 0.27 & 16.87 $\pm$ 12.45 \\
\end{tabular}}
\end{center}
\caption{Performance on BraTS18 dataset. \hl{Percentages in parenthesis indicate the proportion of voxels per class (background: 98.92 \%)}. Scores that were significantly different from $L_{RRW}$ were marked with * ($p$-value $< 0.05$).} \label{table:brats18}
\end{table*}}

{\renewcommand{\arraystretch}{1.2}
\setlength{\tabcolsep}{8pt}
\begin{table*}
\begin{center}
\scriptsize{
\begin{tabular}{l|cc|cc}
& \multicolumn{2}{c}{  Kidney (1.89 \%) } & \multicolumn{2}{c}{ Tumor (0.47 \%) } \\
Loss \hl{function} & Dice & HD (mm.) & Dice & HD (mm.) \\
\hline
$L_{RRW}$ & 0.95 $\pm$ 0.04 & 20.29 $\pm$ 16.64 & 0.57 $\pm$ 0.31 & 56.38 $\pm$ 30.92 \\
\hl{$L_{WCE(RRW)}$} & 0.92 $\pm$ 0.04$^{*}$ & 22.16 $\pm$ 17.18$^{*}$ & 0.51 $\pm$ 0.34$^{*}$ & 56.70 $\pm$ 31.21 \\
$L_{Dice}$ & 0.94 $\pm$ 0.04$^{*}$ & 21.02 $\pm$ 18.44 & 0.53 $\pm$ 0.31 & 84.29 $\pm$ 25.04$^{*}$ \\
$L_{Focal}$ & 0.95 $\pm$ 0.04 & 19.84 $\pm$ 15.73 & 0.51 $\pm$ 0.35$^{*}$ & 41.37 $\pm$ 29.52$^{*}$ \\
$L_{Dice+Boundary}$ & 0.94 $\pm$ 0.05$^{*}$ & 18.62 $\pm$ 16.00 & 0.51 $\pm$ 0.33$^{*}$ & 55.50 $\pm$ 30.48 \\
\end{tabular}}
\end{center}
\caption{Performance on KiTS19 dataset. \hl{Percentages in parenthesis indicate the proportion of voxels per class (background: 97.64 \%)}. Scores that were significantly different from $L_{RRW}$ were marked with * ($p$-value $< 0.05$).} \label{table:kits19}
\end{table*}}

On average, our RRW maps produced the highest Dice coefficients and lowest HDs on ACDC17 dataset (Table \ref{table:acdc17}).
Dice loss yielded significantly worse results in almost all metrics and classes; Focal loss achieved similar or slightly worse results than our RRW maps; \hl{weighted Cross entropy yielded overall worse Dice coefficients and similar HDs}; and Boundary loss also underperformed, except in the HDs of the myocardium class.
In BraTS18 and KiTS19 datasets (Tables \ref{table:brats18} and \ref{table:kits19}), RRW maps produced similar or better Dice coefficient than the other loss functions and, mostly, significantly better HDs than Dice loss.
\hl{Weighted Cross entropy produced similar or worse HDs than the RRW maps}, and Focal and Boundary loss produced overall slightly better HDs, although they were usually not significantly different from our RRW maps.
\hl{Dice coefficients and HDs differed from the competition's top submissions} \cite{bakas2018identifying,bernard2018deep,heller2021state}; \hl{one reason is that the train-validation split utilized in our experiments was more challenging, as the training set was remarkably smaller.
Additionally, in KiTS19 dataset, the underperformance in the tumor class can be further explained by the reduction in the images' resolution that eliminated 75\% of the voxels.}

Our RRW maps achieved, overall, similar or better Dice coefficients and HDs than the other evaluated loss functions in three class-imbalanced datasets.
Notably, as described in Section \ref{sec:regionwisecomparison}, our hyper-parameter-free RRW maps could be tailored to each dataset---strategy similar to hyper-parameter fine-tuning, as in Focal loss---to focus more on certain classes or regions, potentially increasing performance.
However, designing dataset-specific RRW maps was beyond the scope of this paper; instead, we showed that simple RRW maps performed at least as well as other state-of-the-art loss functions.
Dice loss provided segmentations with worse HDs than the other loss functions. This was not surprising as Dice loss considers no pixel importance or prediction confidence, unlike the other loss functions.
\hl{These large HDs also indicate that, despite training on 2D slices of 3D anisotropic medical images has been shown to be beneficial }\cite{isensee2021nnu}\hl{, optimizing Dice loss alone was ineffective.
Regarding Focal loss, although its results were similar to our RRW maps, since Focal loss increases the loss on uncertain predictions (i.e., low softmax values), it can be unsuitable for datasets with noisy labels, which are common in medical image segmentation tasks due to the low inter- and intra-rater agreement.}
\hl{Cross entropy weighted with our RRW maps provided, overall, the lowest Dice coefficients.}
In addition, contrary to Boundary loss, our RRW maps require no hyper-parameter updates during the optimization as the proposed RRW maps with normalized boundary distances are stable to optimize, thus, having no reliance on the training set size.
Furthermore, the validation curves, Figures 1-3 in the Supplementary Material, show that our RRW maps provided faster convergence than Dice, \hl{weighted Cross entropy} and Boundary loss.

\subsection{Evaluation on ACDC17 competition's test set} \label{sec:acdc}
\hl{We evaluated RW loss with our RRW maps on the test set of ACDC17 segmentation competition.
For this, we preprocessed the data and we utilized a nnUNet-based architecture that was as similar as possible to the top submission.
Following Isensee et al. }\cite{isensee2017automatic}\hl{, ACDC17 images were resampled to a common resolution of $1.25 \times 1.25 \times Z_{orig}$ mm. per voxel, and nnUNet was optimized for 300 epochs on batches of 10 2D slices with 48 initial input filters.
nnUNet's default instance normalization layer }\cite{ulyanov2016instance}\hl{ was replaced by batch normalization }\cite{ioffe2015batch}.
\hl{Additionally, since nnUNet could not be trained on patches of size $352 \times 352$ due to a mismatch when concatenating feature maps, we utilized patches of $320 \times 320$ voxels.
The remaining optimization settings, including the learning rate choice and optimizer, were the same as described in Section }\ref{sec:experiments}\hl{.
Finally, data augmentation was performed automatically by nnUNet }\cite{isensee2021nnu}\hl{ and images were normalized to have zero mean and standard deviation of one.}

{\renewcommand{\arraystretch}{1.2}
\setlength{\tabcolsep}{8pt}
\begin{table*} 
\begin{center}
\scriptsize{
\hspace*{-36pt} \begin{tabular}{l|cc|cc|cc|cc|cc|cc}
& \multicolumn{4}{c}{RV} & \multicolumn{4}{c}{M} & \multicolumn{4}{c}{LV} \\
Approach & \multicolumn{2}{c}{Dice} & \multicolumn{2}{c}{HD (mm.)} & \multicolumn{2}{c}{Dice} & \multicolumn{2}{c}{HD (mm.)} & \multicolumn{2}{c}{Dice} & \multicolumn{2}{c}{HD (mm.)} \\
& ED & ES & ED & ES & ED & ES & ED & ES & ED & ES & ED & ES \\
\hline
$L_{RRW}$ (Ours) & 0.95 & 0.89 & 12.0 & 13.9 & 0.91 & 0.92 & 9.3 & 11.8 & 0.97 & 0.93 & 6.2 & 10.5 \\
Isensee et al. \cite{isensee2017automatic} & 0.95 & 0.90 & 10.1 & 12.2 & 0.90 & 0.92 & 8.7 & 8.7 & 0.97 & 0.93 & 7.4 & 6.9 \\
Baumgartner et al. \cite{baumgartner2017exploration} & 0.93 & 0.88 & 12.7 & 14.7 & 0.89 & 0.90 & 8.7 & 10.6 & 0.96 & 0.91 & 6.5 & 9.2 \\
Jang et al. \cite{jang2017automatic} & 0.93 & 0.89 & 12.9 & 11.8 & 0.88 & 0.90 & 9.9 & 8.9 & 0.96 & 0.92 & 7.7 & 7.1 \\
\end{tabular}}
\end{center}
\caption{Performance on ACDC17 competition's test set.} \label{table:acdc17competition}
\end{table*}}

\hl{Table }\ref{table:acdc17competition}\hl{ lists the performance of our method and the top three submissions in ACDC17 segmentation competition }\cite{bernard2018deep}\hl{.
Dice and HD were automatically computed in the online platform separating the images into end diastolic (ED) and end systolic (ES) phases.
RW loss with our RRW maps achieved the same Dice coefficients than the top entry, except in the right ventricle ES, with slightly lower Dice coefficient yet higher than the second entry, and in the myocardium ES, with the highest Dice coefficient by a small margin.
The HDs were, overall, between the first and the second or third entries.
The measurements shown in Table }\ref{table:acdc17competition} \hl{agree with the results of our experiments in Section} \ref{sec:evaluation}\hl{, confirming that the proposed method achieves state-of-the-art performance.}

\subsection{Rectified Region-wise maps convergence} \label{sec:regionwisecomparison}
We quantified the convergence frequency of RW-Boundary maps \hl{(or, equivalently, Boundary loss}, Eq. (\ref{eq:rwboundarymaps})) and our RRW maps (Eq. (\ref{eq:rrwmaps})) in nnUNet, which, in contrast to vanilla UNet, has several architectural components and training settings to facilitate the optimization, such as normalization layers \cite{ulyanov2016instance}, deep supervision \cite{lee2015deeply}, and polynomial learning rate decrease.
We trained 50 models on ACDC17 dataset for 50 epochs, as it was enough to determine whether a model failed to converge, and we calculated the empirical cumulative distribution function (CDF):

\begin{equation}
CDF(d) = \frac{1}{n} \sum_{i=1}^n \mathbbm{1} ( d_i \leq d),
\end{equation}
where $\mathbbm{1}(c)$ is the indicator function taking the value 1 if the condition $c$  is true and 0 if $c$ is false. Intuitively, the CDF indicates the fraction of models that achieved a Dice coefficient of $d$ or lower. 

\begin{figure}[t]
\begin{center}
   \includegraphics[width=0.71\linewidth]{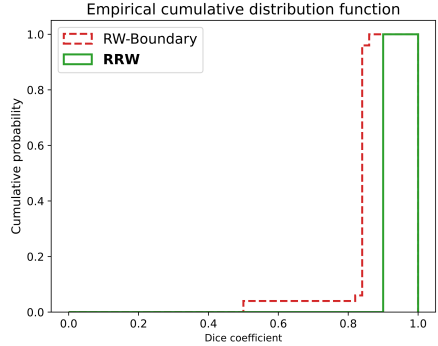}
\end{center}
   \caption{Empirical distribution function illustrating the fraction of models that achieved a  certain Dice coefficient or lower \hl{on ACDC dataset}.}
\label{fig:exp3}
\end{figure}

Figure \ref{fig:exp3} provides empirical evidence that our RRW maps lead to optimization stability, as all models trained with our RRW maps produced Dice coefficients of 0.9 or higher.
In contrast, a fraction of the models optimized with RW-Boundary maps reached a suboptimal local minima due to optimization instability, producing  Dice coefficients around 0.5.
Furthermore, our RRW maps achieved higher Dice coefficients than RW-Boundary maps.

\begin{figure*}[t]
\begin{center}
   \includegraphics[width=0.99\linewidth]{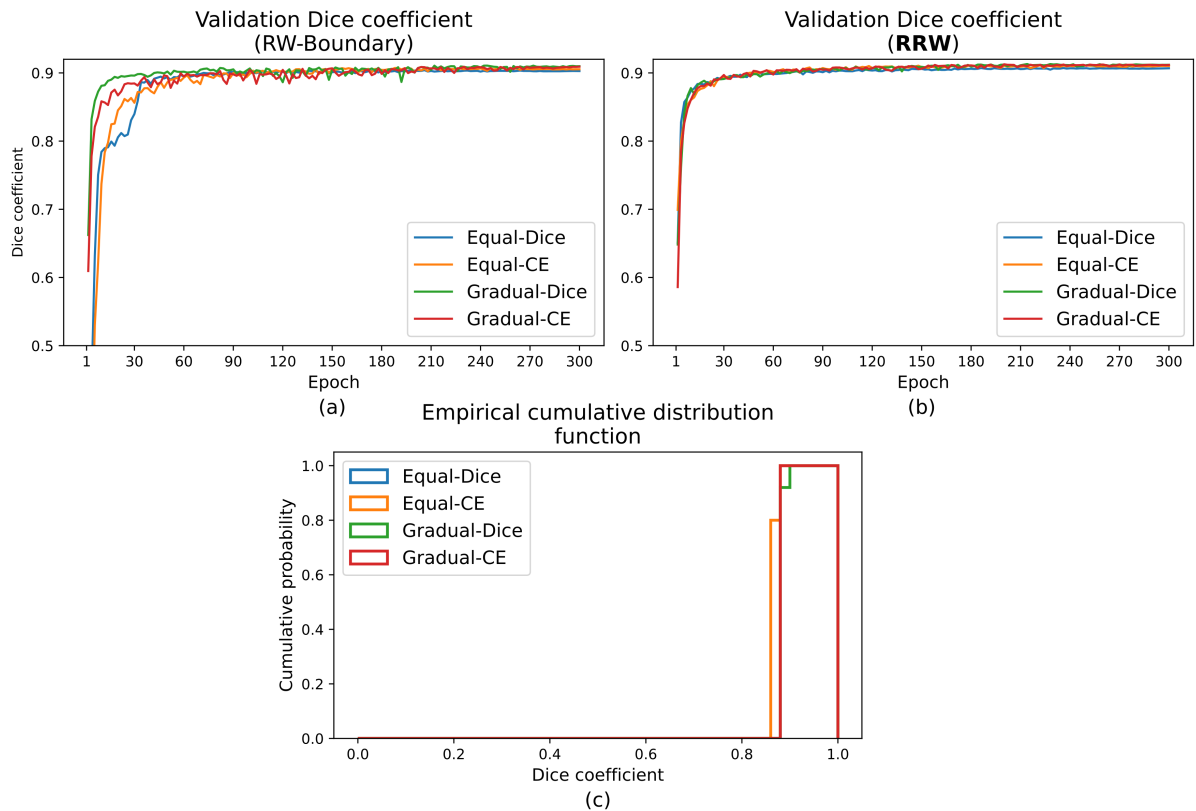}
\end{center}
   \caption{(a-b) Average Dice coefficient during training on ACDC17 dataset while optimizing RW loss with RW-Boundary and RRW maps following different strategies to ensure convergence. (c) Empirical distribution function illustrating the convergence capability of these approaches with RW-Boundary maps.}
\label{fig:exp4}
\end{figure*}

\subsection{Incorporating additional loss functions to facilitate convergence} \label{sec:facilitatingconvergence}

We investigated the impact on the performance and convergence of the two strategies that previous studies utilized to circumvent the optimization instability of RW loss with RW-Boundary maps (Eq. (\ref{eq:rwboundarymaps})).
In particular, we inspected the average Dice coefficient on ACDC17 dataset's validation set during the optimization and, similarly to Section \ref{sec:regionwisecomparison}, we computed the CDF of 50 trained models for each strategy.
These strategies were: 1) combining RW loss with Dice loss: $L_{Dice} + L_{RW}$ \cite{karimi2019reducing} (\textit{equal} contribution); and 2) gradually increasing the influence of RW loss by decreasing $\alpha$ in: $\alpha L_{Dice} + (1 - \alpha) L_{RW}$ \cite{kervadec2018boundary} (\textit{gradual} contribution).
Motivated by the smoothness of Cross entropy's derivative $\frac{\partial L_{RW}}{\partial \boldsymbol{\phi}_{i}} = \boldsymbol{\hat{y}}_i - \boldsymbol{y}_i$, we also studied the use of Cross entropy loss instead of Dice loss\hl{, thus, resulting in four distinct strategies that differed on the extra loss function (Dice or Cross entropy) and the manner in which they were combined with RW loss (equal or gradual contribution)}.
Additionally, although our RRW maps already provide stable optimization, we also evaluated the application of these strategies when optimizing RW loss with the proposed RRW maps (Eq. (\ref{eq:rrwmaps})).

Figure \ref{fig:exp4} (a) shows that, despite averaging across three independently-trained models and three foreground classes, optimizing RW-Boundary maps provided Dice coefficients that fluctuated similarly for the first 210 epochs regardless of the strategy.
At the end of the optimization, the evaluated strategies appear to have converged, providing similar Dice coefficients.
Additionally, \textit{gradual} contribution strategies yielded high Dice coefficients slightly faster than \textit{equal} contribution strategies.
\hl{In contrast, our RRW maps were invariant to the strategy as they equally provided similar performance, oscillations with smaller amplitude between epochs 30-210 (Fig. }\ref{fig:exp4}\hl{ (a-b)), and started to converge at epoch 30---faster than with RW-Boundary maps.
These results corroborate the importance of rectifying RW maps.}

Figure \ref{fig:exp4} (c) illustrates the CDF after optimizing 50 models with RW-Boundary maps.
In contrast to optimizing RW-Boundary maps alone (Section \ref{sec:regionwisecomparison}), incorporating an additional loss function, regardless of the strategy, seemed, in practice, to provide convergence.
In agreement with Figure \ref{fig:exp4} (a), \textit{gradual} contribution strategies were slightly better, increasing the Dice coefficients by a small margin.

\section{Conclusion}
We presented \hl{RW} loss for image segmentation, capable to jointly account for class imbalance and pixel importance. This capability allows to leverage underrepresented classes and to penalize predictions based on pixel location during the optimization.
We showed that other loss functions (Active contour, Hausdorff distance and Boundary losses) can be reformulated as the presented RW loss framework, unveiling their underlying similarities.
Furthermore, our analyses revealed the cause of the optimization instability exhibited by certain special cases of RW loss, such as Boundary loss.
Based on these analyses, we derived a mathematically-grounded principle to provide optimization stability, and we proposed our RRW maps that permit utilizing boundary distance maps while requiring no extra hyper-parameters.

In our experiments, RW loss with our simple hyper-parameter-free RRW maps produced segmentations with Dice coefficients and HDs similar or better than Dice, Focal, \hl{weighted Cross entropy} and Boundary loss functions on three class-imbalanced datasets, achieving state-of-the-art performance.
\hl{Although the RRW maps did not surpass the other losses in every task, RRW can be easily adjusted to consider the particularities of each dataset.
For instance, assigning higher importance to the segmentation boundaries (as in }\cite{caliva2019distance}\hl{) can be useful in images with complex boundaries such as vascular trees.
Additionally, since RW maps can be computed during the optimization, RW maps can be used in areas such as transfer learning or learning with noisy labels.}
We confirmed the optimization instability suffered by Boundary loss, we quantified its frequency, and we empirically demonstrated that our RRW maps are stable to optimize, supporting our theoretical analysis.
Finally, we evaluated multiple strategies to achieve stable optimization with unrectified RW maps and showed that our RRW maps, unlike RW-Boundary maps, barely fluctuated and converged faster.

\section*{Acknowledgements}
The work of J.M. Valverde was funded from the European Union's Horizon 2020 Framework Programme (Marie Skłodowska Curie grant agreement \#740264 (GENOMMED)).
This work has also been supported by the grant \#316258 from Academy of Finland (J. Tohka)

\bibliography{mybibfile}

\newpage
\onecolumn

{\Large Supplementary Material}

\setcounter{section}{0}
\renewcommand\thesection{\Alph{section}}

\section{Full derivation of Region-wise loss gradients}
We derive the gradients of our Region-wise loss $L_{RW} =  \sum_{i=1}^N \boldsymbol{\hat{y}}_i^{\top} \boldsymbol{z}_i$ with respect to the unnormalized prediction of a ConvNet $\boldsymbol{\Phi}$ (i.e., the logits) that multiply every other gradient during optimization with back-propagation. These gradients can be computed by applying the chain rule:

\begin{equation}\label{eq:general}
\frac{\partial L_{RW}}{\partial \phi_{ik}} = \frac{\partial L_{RW}}{\partial \boldsymbol{\hat{y}}_{i}} \frac{\partial \boldsymbol{\hat{y}}_{i}}{\partial \phi_{ik}},
\end{equation}
where $\boldsymbol{\hat{y}}_i$ are the ConvNet predictions at pixel $i$ normalized by the softmax function

\begin{equation}
\boldsymbol{\hat{y}}_i = \sigma(\boldsymbol{\phi}_{i}) =  \frac{\exp(\boldsymbol{\phi}_{i})}{\sum_{l=1}^K e^{\phi_{il}}}.
\end{equation}
First, we calculate the derivative of the softmax function with respect to the logits. The Jacobian of $\boldsymbol{\hat{y}}_i$ is:

\begin{equation}
\frac{\partial \boldsymbol{\hat{y}}_{i}}{\partial \boldsymbol{\phi}_{i}} =\left[\begin{array}{ccc}
\frac{\partial \hat{y}_{i1}}{\partial \phi_{i1}} & \cdots & \frac{\partial \hat{y}_{iK}}{\partial \phi_{i1}}
 \\
\vdots & \ddots & \vdots \\
\frac{\partial \hat{y}_{i1}}{\partial \phi_{iK}}
 & \cdots & \frac{\partial \hat{y}_{iK}}{\partial \phi_{iK}}
\end{array}\right],
\end{equation}
\hl{where, for any $k,l$,}

\begin{equation} \label{eq:generalsoftmaxderivative}
\frac{\partial \hat{y}_{ik}}{\partial \phi_{il}} = \frac{\frac{\partial e^{\phi_{ik}}}{\partial \phi_{il}} \sum_{m=1}^K e^{\phi_{im}} - e^{\phi_{ik}} e^{\phi_{il}} }{ (\sum_{m=1}^K e^{\phi_{im}})^2}.
\end{equation}

\noindent
\hl{Following Eq. }(\ref{eq:generalsoftmaxderivative}\hl{), if $k = l$:}

\begin{equation}\label{eq:der1_1}
\frac{\partial \hat{y}_{ik}}{\partial \phi_{il}} =
\frac{e^{\phi_{ik}}}{\sum_{m=1}^K e^{\phi_{im}}} \frac{\sum_{m=1}^K e^{\phi_{im}} - e^{\phi_{ik}}}{\sum_{m=1}^K e^{\phi_{im}}} = \hat{y}_{ik} (1-\hat{y}_{ik}),
\end{equation}
\hl{and if $k \neq l$:}

\begin{equation}\label{eq:der1_2}
\frac{\partial \hat{y}_{ik}}{\partial \phi_{il}} =
\frac{0 - e^{\phi_{il}} e^{\phi_{ik}}}{(\sum_{m=1}^K e^{\phi_{im}})^2} = - \hat{y}_{ik} \hat{y}_{il}.
\end{equation}
The derivative of our Region-wise loss with respect to the normalized predictions is

\begin{equation}\label{eq:der2}
\frac{\partial L_{RW}}{\partial \boldsymbol{\hat{y}}_{i}} = \boldsymbol{z}_{i}.
\end{equation}
Finally, by plugging Eqs. (\ref{eq:der1_1}-\ref{eq:der1_2}-\ref{eq:der2}) into Eq. (\ref{eq:general}), we have:

\begin{equation}
\begin{split}
\frac{\partial L_{RW}}{\partial \phi_{ik}} & = z_{ik} \hat{y}_{ik} (1 - \hat{y}_{ik}) + \sum_{l = 1; l \neq k}^K z_{il} (-\hat{y}_{ik} \hat{y}_{il}) = z_{ik} \hat{y}_{ik} - z_{ik} \hat{y}_{ik} \hat{y}_{ik} - \sum_{l = 1; l \neq k}^K z_{il} \hat{y}_{ik} \hat{y}_{il} \\
& = z_{ik} \hat{y}_{ik} - \hat{y}_{ik} \sum_{l=1}^K z_{il} \hat{y}_{il} = \hat{y}_{ik} \sum_{l = 1; l \neq k}^K \hat{y}_{il} (z_{ik} - z_{il}).
\end{split}
\end{equation}

\newpage
\section{Region-wise loss generalizes several loss functions in medical image segmentation}

We consider only binary segmentation tasks in this Section, i.e., we assume $K=2$, where the class 1 is the background and the class 2 is the foreground.  Moreover, we let $\Omega$ (without any index) to denote $\Omega_2$ , i.e., the foreground region in the ground-truth.    

\noindent
{\bf Proposition 1} The RW loss with RW map
\begin{equation} \label{eq:aclossorig}
z_{ik}=
    \begin{cases}
        0 & \text{if } i \in \Omega_k \\
        1 & \text{otherwise}
    \end{cases},
\end{equation}
or, equivalently, $\boldsymbol{Z} = \boldsymbol{1} - \boldsymbol{Y}$, where $\boldsymbol{1}$ is the $N \times K$ matrix of ones, is equivalent to AC loss \cite{chen2019learning} without regularization.

\vspace{0.1in}
{\em Proof:} For binary segmentation, AC loss is defined as $L_{AC} = length + \lambda region$ where
\begin{equation} \label{eq:acloss_supl}
    region = \sum_i^N \hat{y}_{i} (c_1 - y_{i})^2 + \sum_i^N (1 - \hat{y}_{i}) (c_2 - y_{i})^2,
\end{equation}
where $c_1=1, c_2=0$ in supervised learning \cite{chen2019learning}, and $y_i = 1$ if $i \in \Omega$ or $y_i = 0$ if $i \notin \Omega$.  Eq. (\ref{eq:acloss_supl}) is equivalent to
\begin{equation} \label{eq:acloss2}
    region = \underbrace{\sum_{i \notin \Omega} \hat{y}_{i2}}_{\text{false positives}} + \underbrace{\sum_{i \in \Omega} \hat{y}_{i1}}_{\text{false negatives}},
\end{equation}
where the first term represents background pixels misclassified as foreground (false positives), and the second term represents foreground pixels misclassified as background (false negatives).
Define $L_{RW \mhyphen AC}$ with a RW map $\boldsymbol{z}_i = [1,0]^T$ if $i \in \Omega$ and $\boldsymbol{z}_i = [0,1]^T$ if $i \notin \Omega$, or, equivalently, $\boldsymbol{Z} = \boldsymbol{1} - \boldsymbol{Y}$, where $\boldsymbol{1}$ is the $N \times K$ matrix of ones. We get 

\begin{equation}
L_{RW \mhyphen AC} = \sum_{i = 1}^N [\hat{y}_{i1}, \hat{y}_{i2}]\boldsymbol{z}_i =  \sum_{i \in \Omega} \hat{y}_{i1} +  \sum_{i \notin \Omega} \hat{y}_{i2}
\end{equation}
and $L_{AC} = length + \lambda L_{RW \mhyphen AC}$. 

\vspace{0.1in}
\noindent
{\bf Proposition 2} The RW loss with RW map
\begin{equation} \label{eq:rwboundarymapsv2}
z_{ik}=
    \begin{cases}
        -{||i - b_{ik}}||_2 & \text{if } i \in \Omega_k \\
        {||i - b_{ik}}||_2 & \text{otherwise}
    \end{cases},
\end{equation}
where $\Omega_k$ is the foreground area of class $k$, and $b_{ik}$ is the closest ground-truth boundary pixel to $i$ in class $k$, is equal to Boundary loss \cite{kervadec2018boundary}.

\vspace{0.1in}
{\em Proof:} For binary segmentation, Boundary loss \cite{kervadec2018boundary} is defined as:

\begin{equation} \label{eq:newboundary}
    \mathcal{L}_{B}(\boldsymbol{\hat{Y}})=\int_{i \in \Omega \cup i \notin \Omega} \phi_{\Omega}(i) \hat{y}_i d i,
\end{equation}
where $\phi_{\Omega}(i)$ denotes the level set representation of the boundary, i.e., $\phi_{\Omega}(i)=-D_{\Omega}(i)$ if $i \in \Omega$ and $\phi_{\Omega}(i)=D_{\Omega}(i)$ otherwise.
Equation (\ref{eq:newboundary}) is equivalent to:

\begin{equation} \label{eq:newboundary2}
\begin{split}
    \mathcal{L}_{B}(\boldsymbol{\hat{Y}}) &= \int_{i \in \Omega} (D_\Omega(i)_1 \hat{y}_{i1} - D_\Omega(i)_2 \hat{y}_{i2} )  d i + \int_{i \notin \Omega} (-D_\Omega(i)_1 \hat{y}_i + D_\Omega(i)_2 \hat{y}_i)d i \\
    &= \sum_{i \in \Omega} [\hat{y}_{i1}, \hat{y}_{i2}] [D_{\Omega}(i)_k, -D_{\Omega}(i)_k]^{\top} + \sum_{i \notin \Omega} [\hat{y}_{i1}, \hat{y}_{i2}] [D_{\Omega}(i)_k, -D_{\Omega}(i)_k]^{\top} .
\end{split}
\end{equation}
Since $D_\Omega(i)_k$ is the distance from pixel $i$ to the nearest pixel in the boundary of class $k$, by defining $z_{ik}$ as in Eq. (\ref{eq:rwboundarymapsv2}) we get:

\begin{equation}
    L_{RW-Boundary} = \mathcal{L}_{B} = \sum_{i=1}^N [\hat{y}_{i1}, \hat{y}_{i2}] [z_{i1}, z_{i2}]^{\top}.
\end{equation}

\vspace{0.1in}
\noindent
{\bf Proposition 3} The modified RW loss $L_{RW^2} = \sum_{i =1}^N[\hat{y}_{i1}^2, \hat{y}_{i2}^2]\boldsymbol{z}_i$ with RW map
\begin{equation} \label{eq:hdloss}
z_{ik}=
    \begin{cases}
        0 & \text{if } i \in \Omega_k \\
        {||i - b_{ik}}||_2^\alpha & \text{otherwise}
    \end{cases},
\end{equation}
where $b_{ik}$ is the closest ground-truth boundary pixel to $i$ in class $k$, is equal to HD loss (one-sided) \cite{karimi2019reducing}.

\vspace{0.1in}
{\em Proof:} For binary segmentation, HD loss (one-sided) \cite{karimi2019reducing} is defined as:

\begin{equation} \label{eq:originalhdloss}
    \operatorname{Loss}_{\mathrm{DT}-\mathrm{OS}}=\frac{1}{N} \sum_{i=1}^N\left((y_i-\hat{y}_i)^{2} d_{y_i}^{\alpha} \right),
\end{equation}
where $d_{y_i}^\alpha$ is the distance to the ground-truth boundary pixel with respect to $i$ to the power of $\alpha$.
Since $y_i - \hat{y}_i \equiv \hat{y}_{i1}$ if $i \in \Omega$ and $y_i - \hat{y}_i \equiv \hat{y}_{i2}$ if $i \notin \Omega$:

\begin{equation}
    \operatorname{Loss}_{\mathrm{DT}-\mathrm{OS}} = \frac{1}{N} \sum_{i \in \Omega} [\hat{y}_{i1}^2, \hat{y}_{i2}^2] [d_{y_{i1}}^{\alpha}+d_{\hat{y}_{i1}}^{\alpha}, 0]^{\top} + \frac{1}{N} \sum_{i \notin \Omega} [\hat{y}_{i1}^2, \hat{y}_{i2}^2] [0, d_{y_{i2}}^{\alpha}+d_{\hat{y}_{i2}}^{\alpha}]^{\top}.
\end{equation}
By defining $z_{ik}$ as in Eq. (\ref{eq:hdloss}), we get:
\begin{equation} \label{eq:hdlossrwloss}
    \operatorname{Loss}_{\mathrm{DT}-\mathrm{OS}} = \frac{1}{N} \sum_{i=1}^N [\hat{y}_{i1}^2, \hat{y}_{i2}^2] [z_{i1}, z_{i2}]^{\top},
\end{equation}
which is equivalent to RW Loss with the softmax predictions squared.

{\bf Note 1.} The square in Eq. (\ref{eq:hdlossrwloss}) comes from using the squared error $(y_i-\hat{y}_i)^{2}$ in Eq. (\ref{eq:originalhdloss}), which penalizes large errors more than the absolute error $|y_i-\hat{y}_i|$.
Although the absolute error is more closely related to the Hausdorff distance and would lead to the original definition of RW loss, Karimi et al. \cite{karimi2019reducing} employed the squared error inspired by the results of Janocha et al. \cite{janocha2017loss}. This necessitated the small modification in the definition of the RW loss to prove the equivalence. 

{\bf Note 2.} Karimi et al. \cite{karimi2019reducing} defined several flavours of their HD loss. We considered here the "one sided" approximation to the original HD loss. The rationale to various approximations in \cite{karimi2019reducing} was that the HD loss is computationally expensive as the distance maps depend on the network parameters and need to re-computed after the parameter updates.  RW loss framework could be generalized to represent also the original HD loss by allowing RW maps $\boldsymbol{Z}$ to depend on the on the network parameters via $\boldsymbol{\hat{Y}}$, but that generalization would lose the simplicity (both computational and conceptual) of the RW loss framework.

\vspace{0.1in}
\noindent
{\bf Proposition 4} The RW loss with RW map
\begin{equation} \label{eq:rwplanevalues}
z_{ik}=
    \begin{cases}
        \alpha & \text{if } i \in \Omega_k \\
        -\beta & \text{otherwise}
    \end{cases},
\end{equation}
where $\alpha, \beta \in \mathbb{R}^N$ is equivalent to $L_{\text {boundary }}$ from Cao et al. \cite{cao2020boundary}.

\vspace{0.1in}
{\em Proof:} The boundary loss function proposed in Cao et al. \cite{cao2020boundary} is defined as

\begin{equation} \label{eq:fourth}
L_{\text {boundary }}=\int_{i \in \Omega \cup i \notin \Omega} \Phi_{\Omega}(i) \hat{y}_i d i
\end{equation}
where $\Phi_{\Omega}(i) = \alpha$ if $i \in \Omega$ and $\Phi_{\Omega}(i) = -\beta$ otherwise, and $\alpha, \beta \in \mathbb{R}^N$. 
Equation (\ref{eq:fourth}) is equivalent to:

\begin{equation}
    \begin{split}
     L_{\text {boundary }} &= \int_{i \in \Omega} -\beta \hat{y}_{i1} + \alpha \hat{y}_{i2} + \int_{i \notin \Omega} \alpha \hat{y}_{i1} - \beta \hat{y}_{i2} \\
     &= \sum_{i \in \Omega} [\hat{y}_{i1}, \hat{y}_{i2}] [-\beta, \alpha]^{\top} + \sum_{i \notin \Omega} [\hat{y}_{i1}, \hat{y}_{i2}] [\alpha, -\beta]^{\top} \\
    &= \sum_{i=1}^{N}[\hat{y}_{i 1}, \hat{y}_{i 2}][z_{i 1}, z_{i 2}]^{\top},
    \end{split}
\end{equation}
where $z_{ik}$ is defined as in Eq. (\ref{eq:rwplanevalues}).

\vspace{0.1in}
\noindent
{\bf Proposition 5} The RW loss with RW map $\boldsymbol{Z} = \boldsymbol{1} - \boldsymbol{Y}$ (as in Eq. (\ref{eq:aclossorig})) is equivalent to the region-based loss in \cite{huang2021dual}.

\vspace{0.1in}
{\em Proof:} The region-based loss in \cite{huang2021dual} is defined as:

\begin{equation}
    RL = \int_{i \in \Omega \cup i \notin \Omega} y_i (y - \hat{y}) + (1 - y_i) \hat{y}.
\end{equation}
Since $y_i = 1$ if $i \in \Omega$ and 0 otherwise

\begin{equation} \label{eq:other}
    RL = \sum_{i \in \Omega} (1-\hat{y}_i) + \sum_{i \notin \Omega} \hat{y}_i.
\end{equation}
For binary segmentation, Eq. (\ref{eq:other}) is equivalent to Eq. (\ref{eq:acloss2}). From here, the proof of equivalence to RW loss is analogous to Proposition 1.

\newpage

\section{Validation curves}
\subsection{ACDC17 Dataset}
\begin{figure*}[htbp]
  \centering
  \includegraphics[width=0.84\linewidth]{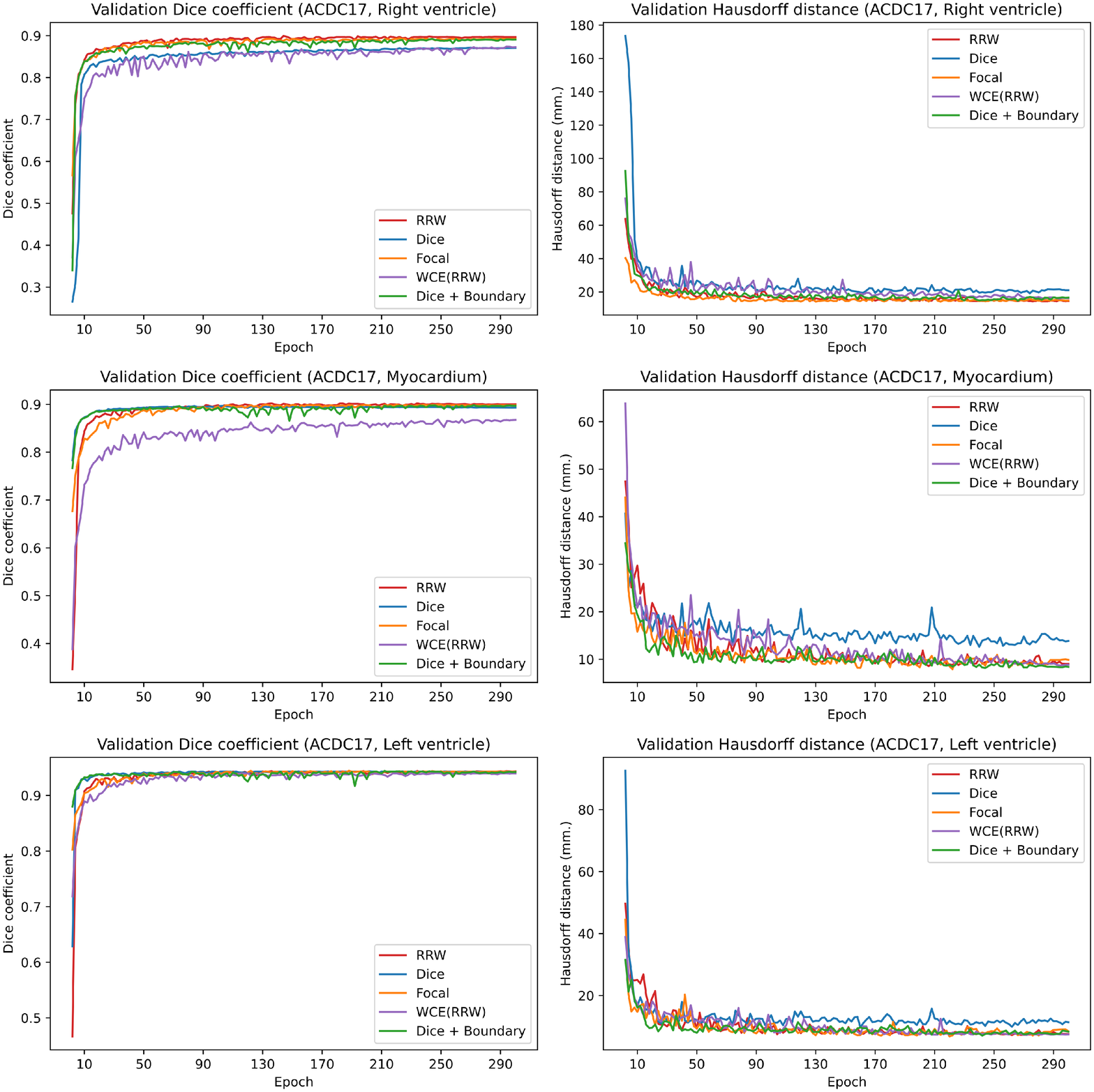}
  \caption{Validation Dice coefficient (left) and Hausdorff distance (right) for all non-background classes in ACDC17 dataset. Values are averages across 3 independent runs.}
\end{figure*}

\newpage
\subsection{BraTS18 Dataset}
\begin{figure*}[htbp]
  \centering
  \includegraphics[width=0.84\linewidth]{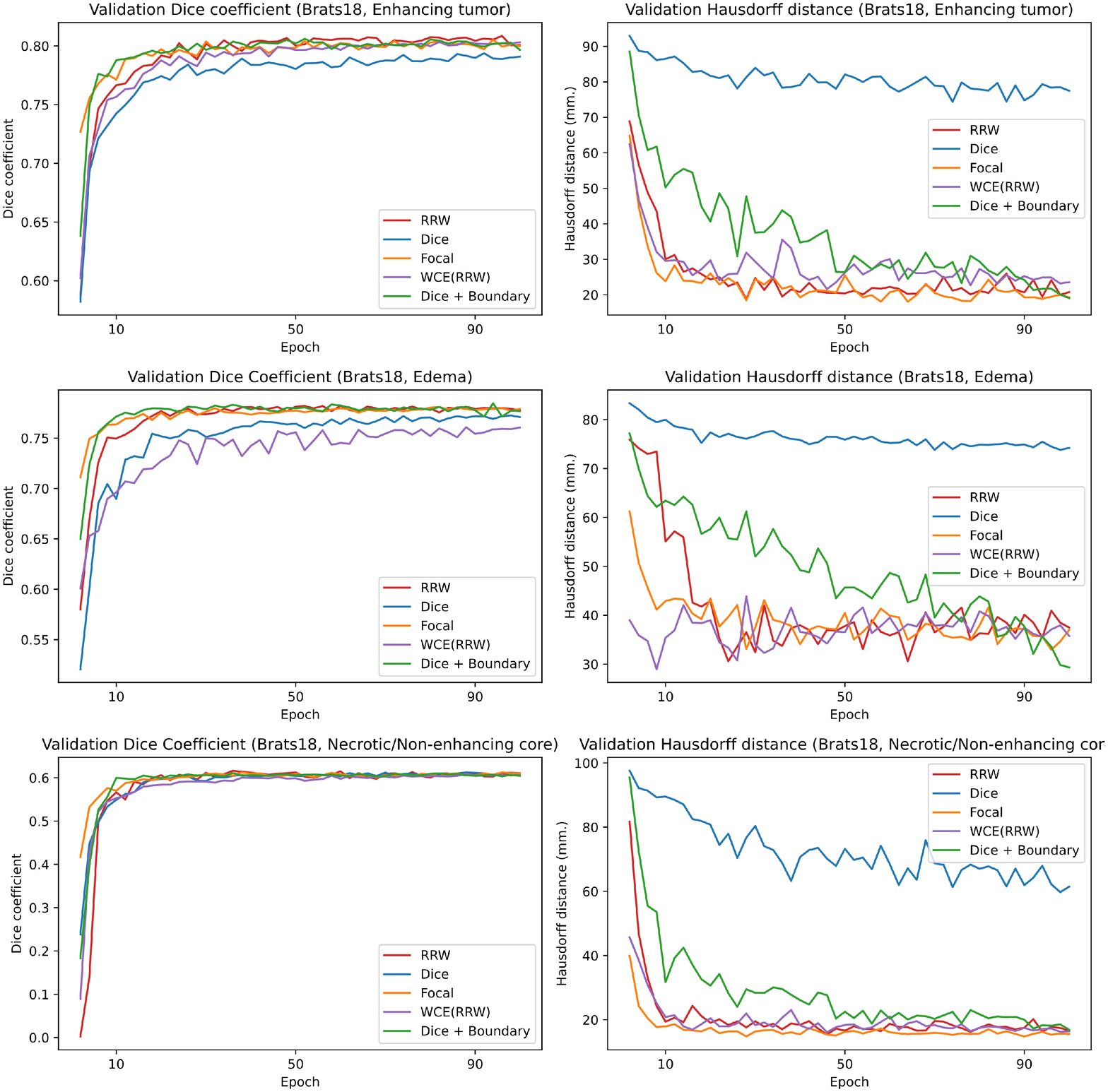}
  \caption{Validation Dice coefficient (left) and Hausdorff distance (right) for all non-background classes in BraTS18 dataset. Values are averages across 3 independent runs.}
\end{figure*}

\newpage
\subsection{KiTS19 Dataset}
\begin{figure*}[htbp]
  \centering
  \includegraphics[width=1\linewidth]{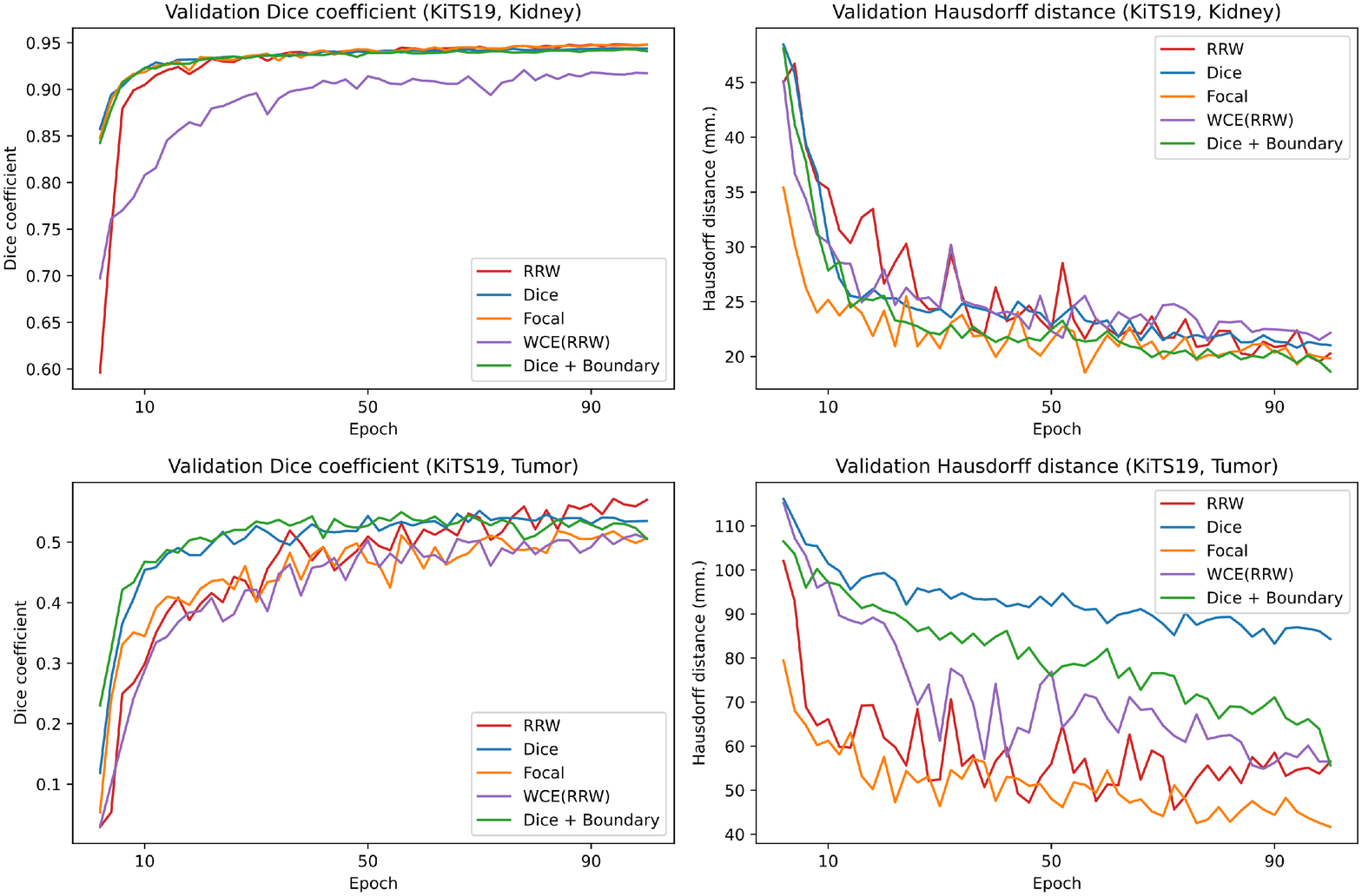}
  \caption{Validation Dice coefficient (left) and Hausdorff distance (right) for all non-background classes in KitTS19 dataset. Values are averages across 3 independent runs.}
\end{figure*}

\end{document}